\documentclass[epj]{svjour}
\usepackage{latexsym}
\usepackage{graphics}
\usepackage{bbm}
\begin{document}
\title{Magnetic properties of the three-band Hubbard model}
\author{Th. Maier \and M. B. Z\"olfl \and
Th. Pruschke \and J. Keller
}                     
%
%
\institute{Institut f\"ur Theoretische Physik I, Universit\"at
Regensburg, Universit\"atsstrasse 31, 93053 Regensburg}
\date{Received: date / Revised version: date}
%
\abstract{
We present magnetic properties of the three-band Hubbard model in
the para- and antiferromagnetic phase on a hypercubic
lattice calculated with the Dynamical Mean-Field Theory (DMFT). To allow for
solutions with broken spin-symmetry we extended the approach to lattices
with AB-like structure. Above a critical sublattice magnetization
$m_d\approx 0.5$ one can observe rich structures in the spectral-functions
similar to the $t$-$J$ model which can be related to the well known
bound states for one hole in the Ne\'{e}l-background. In addition to
the one-particle properties we discuss the static
spin-susceptiblity in the paramagnetic state at the points $\bf{q}=0$
and $\bf{q}=(\pi,\pi,\pi,\cdots)$ for different dopings $\delta$. The
$\delta$-$T$-phase-diagram exhibits an enhanced stability of the
antiferromagnetic state for electron-doped systems in comparison to
hole-doped. This asymmetry in the phase diagram is in qualitative agreement
with experiments for high-T$_c$ materials.
} 
\maketitle
\section{Introduction and Model}
\label{intro}
One of the few undisputed facts about high-T$_c$ materials is that all
undoped high-$T_c$ compounds are insulators with
antiferromagnetic ordering in the $\rm{CuO_2}$-planes at low enough
temperatures \cite{Alm}. Doping of the systems leads
to a strong suppression of the antiferromagnetic order and eventually 
\begin{figure}
\resizebox{0.50\textwidth}{!}{
\includegraphics{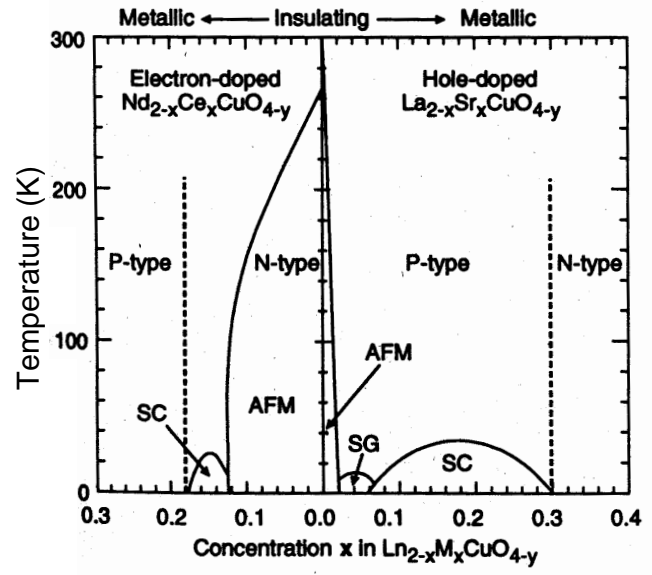}
}
\caption{Experimental phase diagram for high-T$_c$ compounds (taken
from \cite{Alm}).}
\label{expphase}       
\end{figure}
superconductivity sets in. While the explanation of this transition and
the proximity of antiferromagnetism and superconductivity surely is the
most fascinating aspect in the high-T$_c$ compounds, there are a variety of
other peculiarities that call for an explanation. One of these side aspects
is the observation that, although the basic scenario is the same, hole
and electron doped materials show an apparent qualitative difference in that the magnetic
phase in the latter appears to be much more stable for the latter
(cf. Fig. \ref{expphase}).

The physics in the insulating phase is well described by a Heisenberg model since charge fluctuations are suppressed by strong local
correlations and, to a first approximation, one is left with the $\rm Cu$-spin
degrees of freedom only. The exchange
parameter of the resulting effective Heisenberg model is obtained from standard
superexchange processes \cite{Anderson,Zhang},
induced by virtual hopping of a hole from one $\rm Cu$-ion to the
neighbouring one over a nonmagnetic $\rm O$-ion.

At any finite doping one has to consider at least also the charge degrees of
freedom on the copper sites and would then be left with the usual one-band Hubbard
or $t$-$J$ model to describe the interplay between magnetic exchange and itinerancy.
However, this scenario completely neglects the existence of the oxygen sites.
That they are indeed important, at least for the magnetic properties, can be
seen from the following qualitative argument:
Due to the
strong local correlations at the $\rm Cu$-sites, additional doped
holes mainly occupy $\rm O$-sites. The spin of the hole at the $\rm
O$-site induces an effective ferromagnetic interaction \cite{Aha}
between the neighbouring $\rm Cu$-spins, so that the antiferromagnetic
ordering is strongly suppressed with increasing hole-doping. In the case of
electron doping, on the other hand, the additional particle has to go to the
copper sites due to Pauli's principle, which effectively means that free spins
are removed from the system. This ob\-vious\-ly also leads to a suppression of
magnetic order, but in a weaker fashion. Thus, in order to obtain a
more realistic description of the physics of high-T$_C$ compounds one has to take into account 
also the
oxygen degrees of freedom and therefore one-band models like the
standard Hubbard or $t$-$J$ model become inadequate.

The simpliest model which includes both the effect of strong
correlations and the influence of the oxygen sites, is the three-band Hubbard model or Emery model
\cite{Emer}.
The three-band Hubbard Hamiltonian reads
\begin{eqnarray}
H&=&\sum\limits_{i,\sigma}\varepsilon_d d_{i\sigma}^\dagger
d_{i\sigma} + \sum\limits_{j,\sigma}\varepsilon_p p_{j\sigma}^\dagger
p_{j\sigma}\nonumber\\
& +& \sum\limits_{\langle ij \rangle ,\sigma}t_{ij}\left(
d_{i\sigma}^\dagger p_{j\sigma}+h.c.\right) + \sum\limits_{i} U_d\,
n_{i\uparrow}^d n_{i\downarrow}^d\quad\mbox{,}
\label{3bhh}
\end{eqnarray}
where the vacuum is defined as all orbitals in (\ref{3bhh}) filled
with electrons. With this convention $d_{i\sigma}^\dagger(p_{j\sigma}^\dagger)$ 
creates a
hole in a $\rm{Cu}\,3d_{x^2-y^2}$-($\rm{O}\,2p_{x/y}$)-orbital at site
$i(j)$ with spin $\sigma$ and $\varepsilon_d(\varepsilon_p)$ are the
corresponding on-site energies. $t_{ij}=\pm t$ denotes the nearest-neighbour
hopping matrix-element between $\rm Cu$- and $\rm O$-sites and $U_d$
stands for the Coulomb-interaction of two holes, residing at the same
$\rm Cu$-site $i$ with number operators $n_{i\sigma}^d$.

In this paper we want to study magnetic properties of the Hamiltonian (\ref{3bhh}) in
the framework of the DMFT. In this theory the dynamical
renormalizations of the one particle properties become purely local \cite{Metzner,Mueller},
so that they can be obtained from an effective impurity problem
coupled to a self-consistent medium. Due to the additional orbital
degreee of freedom in (\ref{3bhh}) the mapping on the corresponding
effective impurity model is not unique. In order to treat local spin and charge
fluctuations between the $\rm Cu$-site and the surrounding $\rm
O$-sites better than on a mean-field level we use the approach,
developed in ref. \cite{Schmalian}, where a cluster of one $\rm{Cu}\,d$-orbital 
and a normalized bonding combination of the four
surrounding $\rm O$-orbitals is coupled to the effective medium. This
method indeed leads to the anticipated physics on the one-particle
level, namely the formation of a low-lying singlet state -- the Zhang-Rice singlet \cite{Zhang}
-- and at half filling to a charge-transfer insulator
\cite{Schmalian,Zaanen,Zoelfl} in contrast to the Mott-Hubbard scenario for the
one-band model. So far, however, only the paramagnetic state has been
studied in ref. \cite{Schmalian}. In ref. \cite{Doniach} the
metal-insulator (MI) transiton was studied with Quantum Monte Carlo in the context of the
DMFT. There the model was also embedded on a bipartite lattice in order to
take into account the antiferromagnetic symmetry breaking, but in this
article 
the attention was called to the MI-transition.

In order to obtain the phase diagram or look at the behaviour in
the antiferromagnetically ordered state the method has to be extended to
allow for the calculation of susceptibilities or solve the DMFT equations
for lattices with AB-like structure, respectively. A
short review of this generalization together with the technique of
calculating the magnetic susceptibility will be given in the next
section, followed by the discussion of our results in section 3. The
paper will conclude with a summary and outlook in section 4.
\section{Method}
\subsection{The DMFT for the three-band model\label{sec:DMFT}}
Let us begin by summarizing the basic concepts introduced in \cite{Schmalian}
for the three-band Hubbard model. 
In order to construct the DMFT for a Cu-O plaquette, it is convenient to
introduce the Fourier transform of the kinetic part of the
Hamiltonian (\ref{3bhh}) after generalization to $d$ dimensions, which then reads \cite{Schmalian}
\begin{eqnarray}
H=\sum\limits_{{\bf k},\sigma}h_\sigma({\bf k})+\sum\limits_i U_d\,n_{i\uparrow}^d n_{i\downarrow}^d+H_{\rm non-bond}\quad\mbox{,}
\label{FT}
\end{eqnarray}
with
\begin{eqnarray}
h_\sigma({\bf k})&=&\varepsilon_d d_{{\bf
k}\sigma}^\dagger d_{{\bf
k}\sigma} + \varepsilon_p p_{{\bf
k}\sigma}^\dagger p_{{\bf
k}\sigma}\nonumber\\
&+&\sqrt{2d}t\gamma_{\bf k}\left(d_{{\bf
k}\sigma}^\dagger p_{{\bf
k}\sigma}+h.c.\right)\quad\mbox{.}
\end{eqnarray}
Here, $d_{{\bf k}\sigma}$ is the Fourier transform of $d_{i\sigma}$ and
$p_{{\bf k}\sigma}$ is the orthonormalized Fourier transform of the
hybridizing combination of the oxygen orbitals surrounding a given copper site
\cite{Zhang,Schmalian}. The $d-1$ linear combinations, which are
orthogonal to $p_{{\bf k}\sigma}$ were collected into $H_{\rm non-bond}$ and
will be dropped in the following, because they are
decoupled from the remainder of the system. Finally, $\gamma_{\bf k}$ is given
by $\gamma_{\bf k}^2=1-\frac{1}{d}\sum\limits_{\nu=1}^d \cos k_\nu$. In ref. \cite{Schmalian} it was shown, that the rescaling
\begin{equation}
\sqrt{2d}t\gamma_{\bf k}\rightarrow 2t^*\gamma_{\bf
k}^*
\end{equation}
with $\gamma_{\bf
k}^*=\sqrt{1-\frac{\varepsilon_{\bf k}}{\sqrt{2d}}}$, $t^{*}=\mbox{const.}\equiv 1$ and {$\varepsilon_{\bf k}=\sum_{\nu=1}^d
\cos k_\nu$} leads to a nontrivial limit for $d\rightarrow\infty$.
The $d$-Green's function in the DMFT now takes the form
\begin{equation}
G^d_{{\bf k}\sigma}(z)=\left[z-\varepsilon_d-\Sigma^d_\sigma(z)
-\frac{4t^{*^2}-\frac{4t^{*^2}}{\sqrt{2d}}\varepsilon_{\bf k}}{z-\varepsilon_p}\right]^{-1}
\label{Gdofk}
\end{equation}
The new ansatz by Schmalian et al.\ was to write the local $d$-Green's
function to be of the form \cite{Schmalian}
\begin{equation}
G^d_\sigma(z)\begin{array}[t]{l}\displaystyle
=\frac{1}{N}\sum\limits_{\bf k}G^d_{{\bf k}\sigma}(z)\\[5mm]
\displaystyle\stackrel{!}{=}
\Big[z-\varepsilon_d-\Sigma^d_\sigma(z)
-\frac{4t^{*^2}}{z-\varepsilon_p-\Delta_\sigma(z)}\Big]^{-1}\;\;.\end{array}
\label{Gdloc}
\end{equation}
In the DMFT the effective Cu-O cluster lives in a so-called {\em
effective medium}\/, defined via
\begin{equation}
{\cal G}_\sigma(z)^{-1}=G^d_\sigma(z)^{-1}+\Sigma^d_\sigma(z)=
z-\varepsilon_d-\frac{4t^{*2}}{z-\varepsilon_p-\Delta_\sigma(z)}\;\;.
\label{effMedium}
\end{equation}
Note that in this form the coupling to the rest of the system, which
is described by $\Delta_\sigma(z)$ happens
through the $p$-states only. This representation of the local Green's
function is obviously not unique. One could also choose a
representation for the local $d$-Green's function of the form
$G^d_\sigma(z)=\Big[z-\varepsilon_d-\Sigma^d_\sigma(z)-\Delta_\sigma(z)\Big]^{-1}$
where the resonance at $z=\varepsilon_p$ is included in
$\Delta_\sigma(z)$. 
But from a numerical point of view the form (\ref{Gdloc}) is more
convenient because the singularity at $z=\varepsilon_p$ is not
included in the hybridization function which therefore becomes smooth
as a function of frequency. The form (\ref{Gdloc}) of the local
Green's function is just the Dyson equation of an effective impurity
problem consisting of one $d$- and one $p$-orbital, where only the
$p$-orbital hybridizes with the conduction electrons (see eq. 13  in ref. \cite{Schmalian}).  
\subsection{DMFT for the N\'eel state\label{sec:DMFTAB}}
In the antiferromagnetic phase the period of the unit cell of the lattice
is doubled due to the reduced translational symmetry.
Consequently, the volume of the magnetic Brillouin zone (MBZ) is
reduced to one-half of the volume in the paramagnetic state and the
vector ${\bf Q}=(\pi,\pi,\pi,\cdots)$ becomes a reciprocal lattice vector.
These changes in the symmetries of the system can be simply taken into account
by introduction of an AB-sublattice structure \cite{Brand} and reformulating
the theory on an enlarged unit cell containing exactly one A- and one B-site.
Since this procedure does not affect the local two-particle interaction in
the Hamiltonian (\ref{FT}) we will concentrate on the kinetic part for the
derivation of the resulting Hamilton matrix.
We first split the kinetic part of (\ref{FT}) in the following way:
\begin{eqnarray}
H={\sum\limits_{{\bf k}\in{\rm MBZ},\sigma}}\left\{h_\sigma({\bf k})+h_\sigma({\bf k+Q})\right\}
\label{red}
\end{eqnarray}
Note that the $\bf k$-sum runs over ${\bf k}$-points in
the reduced Brillouin zone only! Rewriting (\ref{red}) in terms of the
linear combinations
\begin{equation}
\begin{array}{lcl}
\displaystyle
d_{A/B{\bf k}\sigma}&=&\frac{1}{\sqrt{2}}\left(d_{{\bf k}\sigma}\pm
d_{{\bf k+Q}\sigma}\right)\\[5mm]
\displaystyle
p_{A/B{\bf k}\sigma}&=&\frac{1}{\sqrt{2}}\left(p_{{\bf k}\sigma}\pm
p_{{\bf k+Q}\sigma}\right)\quad\mbox{,}
\end{array}
\end{equation}
acting on the A- or  B-sublattice, respectively, one obtains 
\begin{equation}
H={\sum\limits_{{\bf k}\in{\rm MBZ},\sigma}} \Psi_{{\bf k}\sigma}^\dagger
\underline{\underline{H}}_\sigma ({\bf k}) \Psi_{{\bf k}\sigma}\;\;.
\end{equation}
For simplicity we introduced a spinor notation for the operators
$\Psi_{{\bf k}\sigma}^\dagger=(d_{A{\bf k}\sigma}^\dagger\,p_{A{\bf
k}\sigma}^\dagger\,d_{B{\bf k}\sigma}^\dagger\,p_{B{\bf
k}\sigma}^\dagger)$ and the Hamilton matrix on the sublattices
\begin{equation}
{\underline{\underline{H}}}_\sigma({\bf k})=\left(
\begin{array}{cccc}
\varepsilon_d+\Sigma_A^\sigma&\Pi^+_{{\bf k}}&0&\Pi^-_{{\bf k}}\\
\Pi^+_{{\bf k}}&\varepsilon_p&\Pi^-_{{\bf k}}&0\\
0&\Pi^-_{{\bf k}}&\varepsilon_d+\Sigma_B^\sigma&\Pi^+_{{\bf k}}\\
\Pi^-_{{\bf k}}&0&\Pi^+_{{\bf k}}&\varepsilon_p
\end{array}\right)\quad\mbox{.}
\end{equation}
The quantities $\Sigma_{A/B}^\sigma$ denote the local self-energies due to the
two particle term in (\ref{FT}) on A/B-sublattice sites, which are
different in the antiferromagnetic state. Furthermore $\Pi^\pm_{{\bf k}}=\frac{2t^*}{\sqrt{2}}\left(\gamma_{\bf
k}^*\pm\gamma_{\bf k+Q}^*\right)$. For the $d$-components
of the Green's function matrix we finally obtain
\begin{equation}
\underline{\underline{G}}_d^\sigma({\bf
  k},z)=C_{{\bf k}\sigma}\left(
\begin{array}{cc}
\xi_B^\sigma&-\frac{\displaystyle 4t^{*^2}\varepsilon_{\bf
    k}}{\displaystyle\sqrt{2d}\xi_p}\\
-\frac{\displaystyle 4t^{*^2}\varepsilon_{\bf
k}}{\displaystyle\sqrt{2d}\xi_p}&\xi_A^\sigma
\end{array}\right)
\label{d_alone}
\end{equation}
with $C_{{\bf
k}\sigma}=\left[\xi_A^\sigma\xi_B^\sigma-\left(\frac{4t^{*^2}\varepsilon_{\bf
k}}{\sqrt{2d}\xi_p}\right)^2\right]^{-1}$, $\xi_p=z+\mu-\varepsilon_p$
and $\xi_{A/B}^\sigma=z+\mu-\varepsilon_d-\Sigma_{A/B}^\sigma-\frac{4t^{*^2}}{\xi_p}$. The
local Green's function is obtained by taking one of the diagonal
elements and summing over $\bf k$ with the
result
\begin{equation}
G_{d,A/B}^\sigma (z)=\int\limits_{-\infty}^\infty
d\varepsilon\,\rho_o(\varepsilon)\frac{\xi^\sigma_{B/A}}{\xi_A^\sigma
\xi_B^\sigma-\left(\frac{{\displaystyle
4t^{*^2}\varepsilon}}{{\displaystyle \xi_p}}\right)^2}\quad\mbox{,}
\label{G_d_local}
\end{equation}
where the density of states $\rho_o(\varepsilon)$ corresponding to the dispersion
$\varepsilon_{\bf k}$ was introduced.
In the paramagnetic state, where $\xi^\sigma_A=\xi^\sigma_B$, one
immediately recovers the result of section \ref{sec:DMFT}.

In the antiferromagnetic state it is sufficient to perform the calculations
for the A-sublattice only due to the additional symmetry \cite{Brand}
$G_{d,A}^\sigma=G_{d,B}^{\bar{\sigma}}$ and use the spin-index for
book-keeping.
The actual calculation is now a straightforward extension of the method
used in ref. \cite{Schmalian} for the paramagnetic phase. The local nature
of the selfenergies allows the mapping of the lattice problem on an
effective impurity-problem, consisting of a $d$-orbital and the
orthonormalized hybridizing combination of the four surrounding
$p$-orbitals, coupled to the effective medium, which is described
by the propagator (\ref{effMedium}) and has to be determined selfconsistently. Again, the coupling to the
surrounding clusters is assumed to happen through the $p$-states only.

The remaining local
problem is solved with the resolvent method \cite{Kei,Bickers} and an
extended version of the so called Non Crossing Approximation (NCA) \cite{Bickers},
where the 16 local eigenstates of the impurity are coupled through the
hybridization-function $\Delta^\sigma(z)$ \cite{Schmalian}.
\subsection{On the calculation of the magnetic susceptibility\label{sec:MagSusz
}}
On the one-particle level one can obtain magnetic properties by
applying a staggered magnetic field and calculating the sublattice
magnetization. This technique is very tedious so that we used another
method for calculating the magnetic phase diagram.

In addition to the one-particle properties the DMFT also allows to
calculate two-particle correlation functions, e.g.\ the magnetic
susceptibility consistently. In analogy
to the one-particle case the two-particle self energy becomes purely
local in the limit {\nolinebreak $d\rightarrow\infty$} \cite{Brand,Jarr}.
This enables us to extract the two-particle self-energy from the effective
local problem \cite{Jarr,MJ,ThP} and use it to determine the two-particle
correlation function for the lattice.
Since the local two-particle propagator is a function of three frequencies in
the most general case the algorithm works best for Matsubara frequencies,
because all quantities can be represented as matrices in this case.
For details of the method see e.g.\ ref.\ \cite{ThP}.

The choice of the cluster as effective impurity and the finite value of
$U_d$ results in 16 local eigenstates. This leads to a huge number of diagrams for the local two-particle propagator, which have to be 
calculated as functions of three frequencies and summed up
numerically. Although the problem of generating the correct diagrams for the local two-particle
propagator can be automated and handled by the computer, the remaining
numerical task is still formidable and restrict our calculations to
the evaluation of the static
susceptibility for the time being. Nevertheless, the study of the dynamical
susceptibility is in principle also possible \cite{ThPII} and will
be the subject of a forthcoming publication.
\section{Results\label{sec:RESULTS}}
\subsection{Susceptibility and phase diagram\label{sec:SUS}}
\begin{figure}
\resizebox{0.50\textwidth}{!}{
\includegraphics{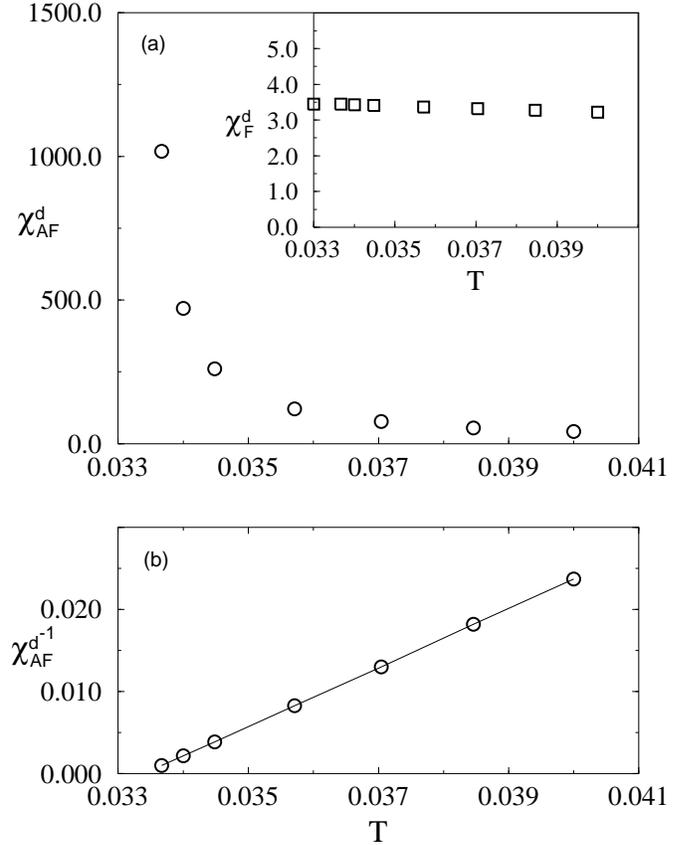}
}
\caption{Homogeneous ($\chi^d_F$) and staggered ($\chi^d_{AF}$)
susceptibilities (a) and inverse staggered susceptibility (b) for 10\% hole-doping and parameters as given in the
text.}
\label{sus}       
\end{figure}

\begin{figure}
\resizebox{0.50\textwidth}{!}{
\includegraphics{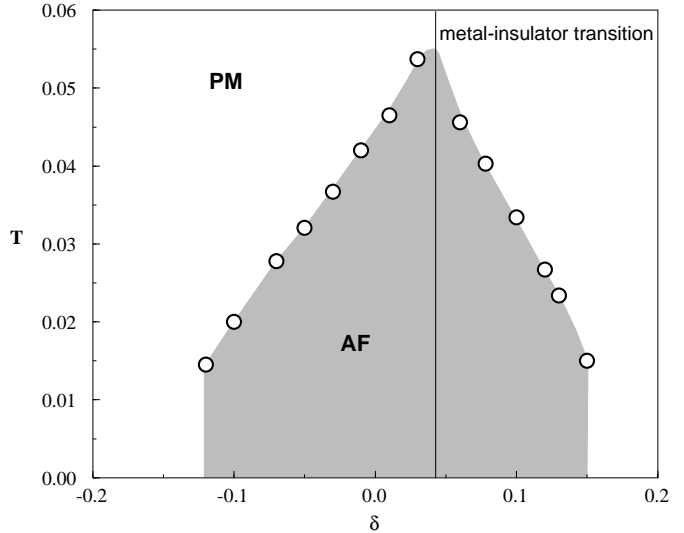}
}
\caption{Magnetic phase-diagram for the 3-band Hubbard model on a
hypercubic lattice for $U_d=2\Delta=7t^*$.}
\label{phase}       
\end{figure}

Let us start the discussion of our results with the magnetic susceptibility. We calculated the static magnetic susceptibility of the $\rm Cu$-spins
 in the paramagnetic phase at the points ${\bf q}=0$ and ${\bf q}={\bf
Q}$, which give the homogeneous and staggered susceptibilities, respectively. For the parameters of the
three-band Hubbard model we have chosen $U_d=2\Delta=7t^*$, where the
charge transfer gap $\Delta$ is defined by
$\Delta=\varepsilon_p-\varepsilon_d$. {\nolinebreak Fig. \ref{sus}a}
 shows typical 
results for these two susceptibilities as a function of temperature
T for a hole-doping $\delta=n_d+n_p-1=0.1$.
For the above choice of $U_d$ and $\Delta$ we observe a finite and
slowly varying ferromagnetic susceptibility $\chi_F^d(T,\delta)$,
which does not show any tendency towards an instability in the calculated
region of temperatures and dopings. The antiferromagnetic
susceptibility $\chi_{AF}^d(T,\delta)$, on the other hand, varies
strongly as a function of temperature and diverges at a finite temperature
$T=T_N$. Fig. \ref{sus}b shows the inverse staggered susceptibility
for the same parameters. As expected we find the linear variation of $\chi_{AF}^{-1}$, which is
typical for a mean-field
theory. By calculating the inverse susceptibility for different
dopings $\delta$ we obtain the $\delta$-$T$-phase-diagram, shown in
{Fig. \ref{phase}}.
Note, that half-filling ($\delta=0$) does not coincide with the MI-transition. This is so because for the used parameter
values of $U_d$ and $\Delta$ the metal-insulator-transition is
shifted towards larger hole-filling values $n>1$. Obviously the
highest value for the Ne\'{e}l temperature is achieved at the
metal-insulator-transition and not at half-filling. As already
observed for the one band model \cite{MJ,ThP} the
antiferromagnetic phase is strongly suppressed upon doping. However,
in contrast to the former case one re\-co\-gnizes a pronounced
asymmetry in the Ne\'{e}l temperature with respect to hole-
and electron-doping. This stronger sensistivity of
the antiferromagnetic ordered state in the case of hole doping
compared to electron doping is qualitatively in good agreement with
experiments (see Fig. \ref{expphase}) and is a direct consequence of the oxygen degrees
of freedom. The present large $D$ treatment however neglects short
range order phenomena so that the argumentation of ref. \cite{Aha}
concerning the frustration of the AF-order by hole doping does not
hold on this level. But nevertheless the oxygen degrees of freedom cause a
particle hole asymmetry in the half filled three band model which
yields the calculated asymmetry in the phase diagram in this large $D$ approach.   

In our calculations the ordered phase is stable up to $\delta\approx
0.18$ for low enough temperatures (cf. Fig. \ref{phase}). Experiments and other
theoretical calculations show much stronger
suppression of the antiferromagnetic ordering with doping \cite{Alm}. The
tendency to overestimate the magnetic phase boundary is typical for
mean field theories, since they completely neglect fluctuations, which
strongly renormalize transition temperatures.
\subsection{Spectral functions in the ordered phase\label{sec:SPEC}}
With the generalized equation (\ref{G_d_local}) it is also possible to
perform calculations in the antiferromagnetic phase. To allow for
solutions with  finite sublattice magnetization\linebreak {\nolinebreak $m_d=\left|\left\langle
n_{d\uparrow}^{A/B}-n_{d\downarrow}^{A/B}\right\rangle\right|$} we apply a
small symmetry-brea\-king staggered magnetic field $h({\bf r}_i)=h
e^{i{\bf Q}\cdot{\bf r}_i}$ in $z$-direction at the beginning of our iteration
procedure, which is turned off after a few iterations.

\begin{figure}
\resizebox{0.50\textwidth}{!}{
\includegraphics{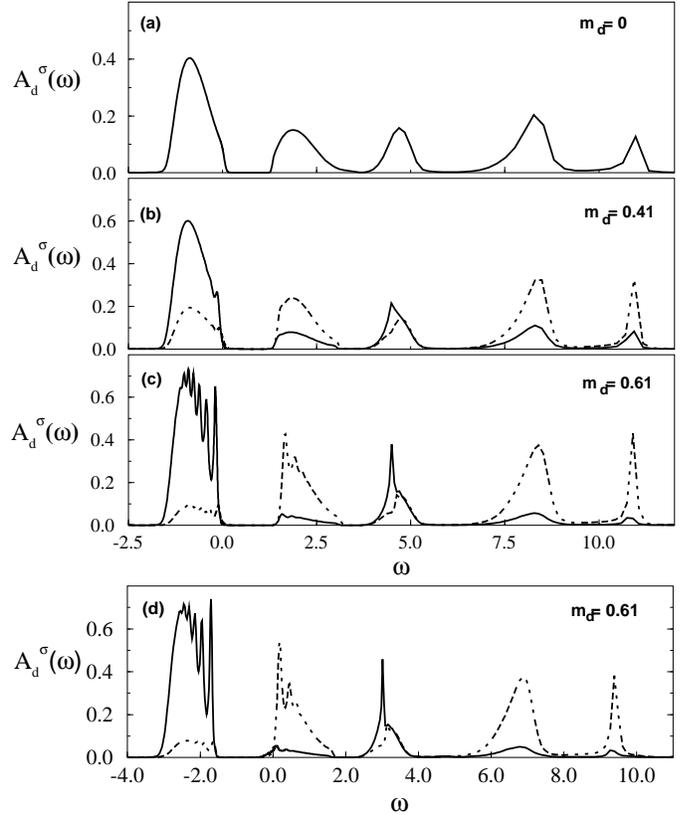}
}
\caption{$d$-spectral function for the majority (full curves) and minority spin 
(dashed curves) for $U_d=2\Delta=7t^*$, fixed doping
$\delta=0.015$ (below the MI-transition (cf. Fig.\ref{phase})) for various temperatures $\beta=18/t^*$ (a),
$\beta=20/t^*$ (b) and $\beta=22/t^*$ (c) and for $\delta=0.063$
(above the MI-transition) and $\beta=22/t^*$ (d)}
\label{af_T}       
\end{figure}

In the following we concentrate on the $d$-part of the spectrum $A_d^\sigma
(\omega)=-\frac{1}{\pi}{\rm Im}G_{d,A}^\sigma
(\omega+i\delta)$ on the A-sublattice, since the $p$-part shows
exactly the same features as the $d$-part, only with different spectral
weights for the various bands. Fig. \ref{af_T}a,b,c shows the typical behaviour as the
temperature $T$ is lowered for fixed values of the parameters
{$U_d=2\Delta=7t^*$} at finite doping $\delta=0.015$.

For $\beta=18/t^*$ the system is in the paramagnetic state
(cf. Fig. \ref{af_T}a), where the general equation (\ref{G_d_local})
reduces to the form (\ref{Gdloc}). We find the same result for the
$d$-part of the spectrum as in
ref. \cite{Schmalian}, where a detailed  discussion of the various bands concerning
their doping dependence, transfer of spectral weight and the evolution
of coherent quasiparticles near the Fermi-energy can be found. Let us
just briefly mention the important low energy parts, namely the
so-called lower Hubbard band at $\omega<0$ in the spectrum of
Fig. \ref{af_T}, which has mainly $d$-character, and the Zhang-Rice
band right above the gap, which is generated by the singlet
combination of the $p$-and $d$-states on one plaquette
\cite{Schmalian}. Note that $\delta >0$ although the chemical
potential is located in the lower Hubbard band. Therefore the MI-transition
occurs at larger filling values as already mentioned in section \ref{sec:SUS}. 
With decreasing
temperature the system enters the antiferromagnetic phase and the
spectral functions of up and down spin become inequivalent (cf. Fig. \ref{af_T}b), yielding a finite sublattice magnetization
$m_d$. Note that the major effect is a transfer of spectral weight from the minority
spin to the majority spin. In addition the peaks in the spectra are slightly shifted
in energy with respect to each other. This effect can be ascribed to
an internal molecular field, generated by the finite sublattice magnetization. Therefore, measuring the energy shift one can calculate the internal
molecular field and from this the exchange parameter $J$ of a
corresponding tJ-model. 
For still lower temperatures the sublattice magnetization increases
and above a value of $m_d\approx 0.5$ a pronounced multipeak
structure is evolving (see Fig. \ref{af_T}c). Fig. \ref{af_T}d shows the
spectral function for the same system parameters and sublattice
magnetization as in Fig. \ref{af_T}c but with the chemical potential
right above the gap. Also in this regime the same multipeak structure
occurs and the spectral function shows little difference to Fig. \ref{af_T}c. Only the peaks next to the chemical potential have more
spectral weight compared with Fig. \ref{af_T}c.

These regular resonances were pre\-vious\-ly found
in DMFT cal\-cu\-lations of the tJ-model \cite{Obermeier}. There the
multiple peaks could be
related to bound states of one single hole in the Ne\'{e}l background. In ref. \cite{Strack} it was shown, that this special problem can be
 solved exactly for $d=\infty$ and $T=0$ within the tJ-model. The most important
physical aspect is, that the moving hole feels a binding potential
proportional to $J$,
growing linearly with the distance from its starting point due to the
breaking of antiferromagnetic bonds during its motion \cite{Strack,Obermeier}. This linear potential leads to a sequence of discrete poles at frequencies \cite{Strack} 
\begin{equation}
\omega_n=-2\hat{t}-\frac{J^*}{2}-a_n\hat{t}\left(\frac{J^*}{2\hat{t}}\right)^\frac{2}{3}
\label{Ai}
\end{equation}
as spectrum for the one particle excitations. Here, the $a_n$ denote
the zeros of the Airy function $Ai\left(4\hat{t}/J^*\right)$, and the
renormalized parameters $\hat{t}$ and $J^*$ are given by
$\hat{t}=t\sqrt{2d}$, $J^*=J2d$. These exact results can be compared
directly to the resonances, found in the DMFT calculations for the
tJ-model \cite{Obermeier}. Since the model used in our calculations is
fundamentally different from the tJ-model and also from the one-band
Hubbard model, the relevance of this physically intuitive picture to
it and especially the proper choice of the parameters for an effective
tJ-model to describe the low energy properties is not
clear a priori. The approach chosen here is to fix the hopping to $t^2/\Delta$,
which reproduces the free bandwidth. In our case we determine an
effective exchange interaction $J^*$ from the energy shift $\Delta
E=J^* m_d$ of the bands. Note that this is just the energy
shift of the spin-up and spin-down bands of a corresponding tJ-model,
treated on a mean-field level \cite{Mueller,Obermeier,Itz}. Another
possibilty to obtain the exchange integral $J$ is to use the result
of a Schrieffer-Wolf transformation of the 3-band Hubbard model, see e.g. \cite{Zaanen}. However this transformation holds only for large values of $U_d$ and $
\Delta$, so that we do not expect this procedure to give a meaningful
result for our parameter values. 

\begin{figure}
\resizebox{0.50\textwidth}{!}{
\hspace{1cm}\includegraphics{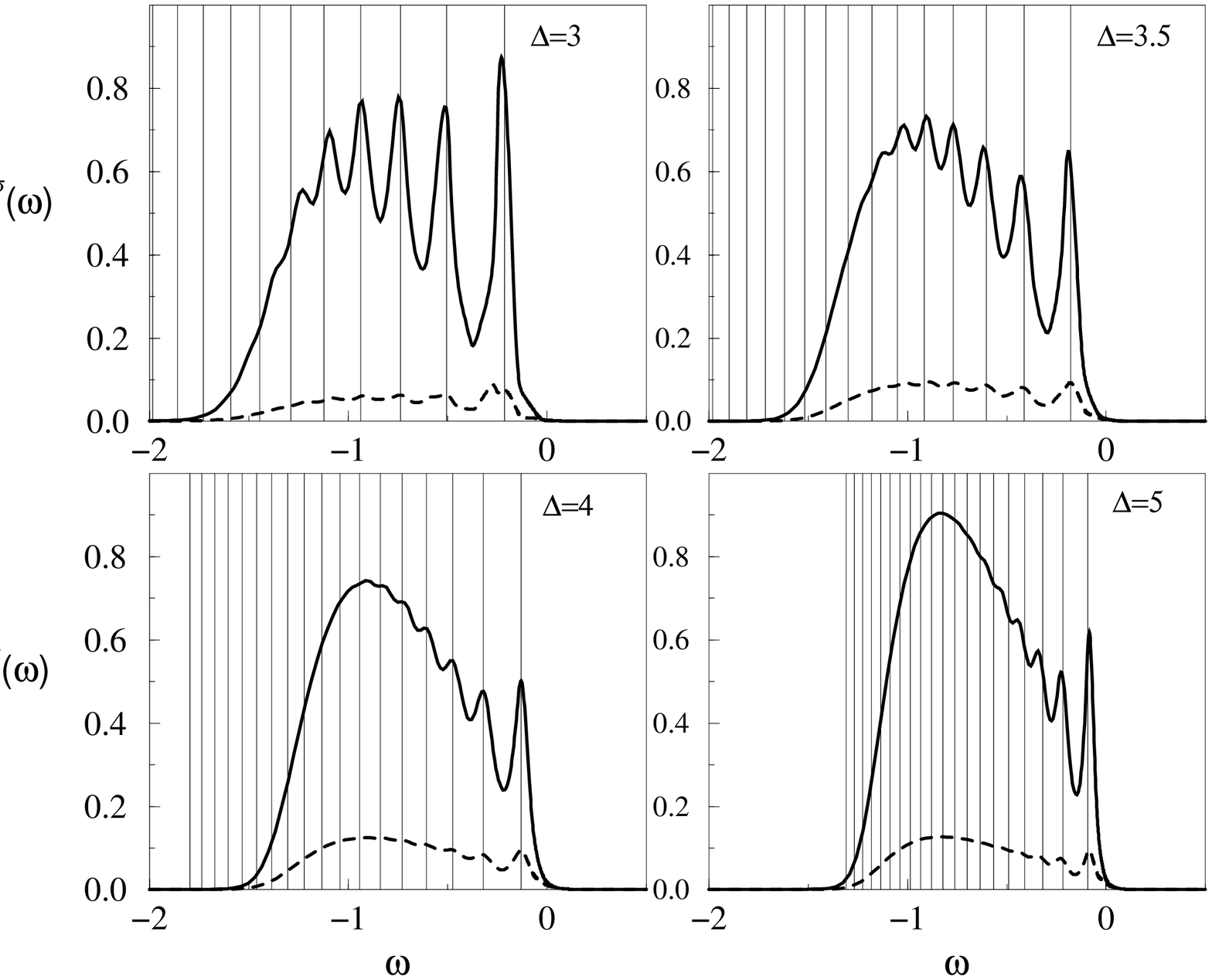}
}
\caption{Comparison of the low energy part of the $d$-spectrum with the exact results (vertical lines), given by equation (\ref{Ai}).}
\label{comp}       
\end{figure}

Fixing the paramters in eq. (\ref{Ai}) as discussed above, we can
indeed directly compare our results with the discrete spectrum (\ref{Ai}). Fig. \ref{comp} shows some examples for the fit of the low energy part of the $d$-spectrum $A_d^\sigma(\omega)$ by the discrete spectrum (\ref{Ai}) at fixed doping $\delta=0.015$ and sublattice magnetization $m_d=0.60$ for various parameters $U_d=2\Delta$.
Note, that the energy scales of up- and down-spin in Fig. \ref{comp} are already shifted
by $\pm \Delta E/2$ respectively, so that the resonances of majority- and
minority-spin bands coincide.      
We find quite good agreement with the distance of the peak
positions. The broadening is expected to result from finite
temperature, sublattice magnetization and doping effects
\cite{Obermeier}. This means that the tJ-model with proper choice of the parameters $t$ and $J$ seems to
reproduce the low energy one-particle dynamics of the three-band Hubbard model in
$d=\infty$ correctly, even in the antiferromagnetic state. In
addition, the basic physical picture for the multipeak structures
observed for low temperatures appears to be the same as in the simple
one-band models.

\begin{figure}
\resizebox{0.50\textwidth}{!}{
\includegraphics{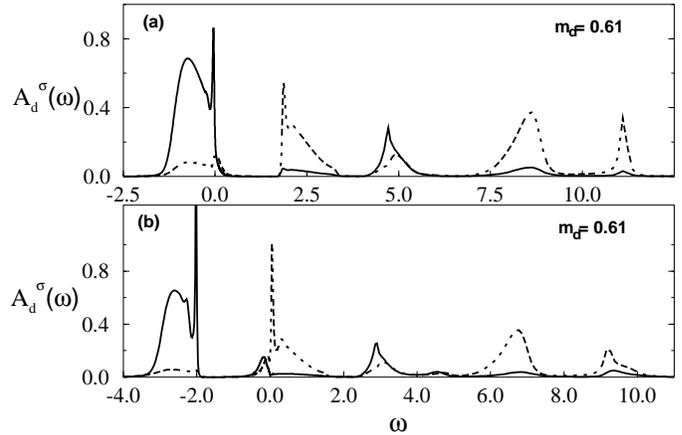}
}
\caption{$d$-spectral function for the majority (full curves) and minority spin 
(dashed curves) for $U_d=2\Delta=7t^*$ and $\beta=50/t^*$ at
$\delta=-0.08$ (a) and $\delta=0.13$ (b).}
\label{af_doped}       
\end{figure}

In order to gain more insight in the effect of doping on these
multipeak structure we investigated the spectrum at larger doping far
away from the MI-transition. Fig. \ref{af_doped} shows the results for the
$d$-part of the spectrum for the same system parameters and sublattice magnetization as in
Fig. \ref{af_T}c,d at $\beta=50/t^*$ but at larger doping
$\delta=-0.08$ (a) and $\delta=0.13$ (b).
In the electron (Fig. \ref{af_doped}a) as well as in the hole doped
regime (Fig \ref{af_doped}b) only the resonances
next to the chemical potential survive. Due to the larger doping there
are more electrons/holes in the system whose paths can intersect and
restore the antiferromagnetic background. Therefore the
electrons/holes become more mobile and the resonances at higher energies are
washed out.    

In finite dimensions the string picture for one hole in the
antiferromagnetic background no longer holds and is correct only up to
order $\frac{1}{d^2}$ \cite{Strack} due to the possibility of paths
which intersect and touch themselves \cite{Trug}. Second, fluctuations
become more important which can restore the antiferromagnetic
background. Thus in low dimensions we expect that the multipeak structure
at finite doping will dissappear.
\section{Summary\label{SEC:SUM}}
In this paper we presented results for the magnetic properties of the
three-band Hubbard model in the limit of high spacial
dimensions. These were obtained in the framework of the Dynamical Mean
Field Theory, which enabled us to calculate the one particle spectrum
as well as two particle correlation functions, namely the magnetic
susceptibility. From this we evaluated the $\delta$-$T$-phase diagram,
which shows strong suppression of the antiferromagnetic state upon
doping. In contrast to one-band models the ordered state is found to
be more sensitive upon doping in the case of hole doping in comparison
to electron doping. This asymmetric behaviour is qualitatively in good
agreement with experiments. The spectral function for single particle
excitations in the antiferromagnetic phase shows pronounced features
above a sublattice magnetization $m_d\approx 0.5$. These
structures are similar to those found in the tJ-model for the special
case of one single hole, moving in the Ne\'{e}l background and can be understood by the binding of one hole in a string potential. A quantitive fit of the spectral
functions by the exact results for the special case for the tJ-model
shows quite good
agreement, so that the tJ-model seems to reproduce the correct physics
of the three-band Hubbard model as long as one is only
interrested in the low energy one-particle physics.

In low dimensions fluctuations become more important which will
destroy the multiple peaks found in the spectral function for
$d=\infty$. Thus these peaks have not yet been observed in
experiments.


%
%

\end{document}